\documentclass[fleqn,usenatbib]{mnras}

\usepackage[T1]{fontenc}
\usepackage{ae,aecompl}

\usepackage{fontawesome}
\usepackage{orcidlink}

\usepackage{graphicx}	
\graphicspath{{imgs/}}
\usepackage{amsmath}	
\usepackage{amssymb}	
\usepackage{xcolor}
\usepackage{color, colortbl}
\usepackage[utf8]{inputenc} 

\newcommand{\Msun}{\text{M}_{\odot}}	    
\newcommand{\fb}{FB15}                      
\newcommand{\fbb}{FB30}                     

\usepackage{newtxtext,newtxmath}

\title[From HI to dark matter with EMBER-2]{Unveiling the dark Universe with HI and EMBER-2}

\author[Bernardini et al.]{
Mauro Bernardini$^{\orcidlink{0000-0002-2930-9509}}$,$^{1}$\thanks{\href{mailto:mauro.bernardini@uzh.ch}{\url{mauro.bernardini@uzh.ch}}}
Robert Feldmann$^{\orcidlink{0000-0002-1109-1919}}$,$^{1}$
Daniel Anglés-Alcázar$^{\orcidlink{0000-0001-5769-4945}}$,$^{2}$ \newauthor \thanks{Following authors are listed in alphabetical order.}
Philipp Denzel$^{\orcidlink{0000-0003-0126-0659}}$,$^{3}$
and Jindra Gensior $^{\orcidlink{0000-0001-6119-9883}}$$^{4}$
\newauthor
\\
$^{1}$Department of Astrophysics, Universität Zürich, Winterthurerstrasse 190, 8057, Zurich, Switzerland\\
$^{2}$Department of Physics, University of Connecticut, 196 Auditorium Road, U-3046, Storrs, CT 06269-3046, USA\\ 
$^{3}$Centre for Artificial Intelligence, Zurich University of Applied Sciences ZHAW, Technikumstrasse 71, Winterthur 8400, Switzerland\\
$^{4}$Institute for Astronomy, University of Edinburgh, Blackford Hill, Edinburgh, EH9 3HJ, UK\\ 
}
\date{Accepted XXX. Received YYY; in original form ZZZ}

\pubyear{2025}

\begin{document}
\label{firstpage}
\pagerange{\pageref{firstpage}--\pageref{lastpage}}
\maketitle

\begin{abstract}
Next-generation radio telescopes will provide unprecedented data volumes of the neutral hydrogen (HI) distribution across cosmic time.
The spatial and kinematic distribution of HI is a biased tracer of the underlying matter field, and as such contains information on the distribution of dark matter over a wide range of scales. Extracting dark matter properties from HI, however, is non-trivial because baryonic processes linked to galaxy formation significantly modify the HI distribution.
Additionally, methods that use empirical relations, often calibrated via numerical simulations, do not use the full field-level information to model the complex relation between HI and dark matter.
We use the recently introduced EMBER-2 model to directly predict dark matter distributions from HI tracers over a wide redshift range, $z=0-6$.
After training on cosmological galaxy formation simulations run with FIRE-2, our method accurately recovers key statistics, including dark matter mass fractions, surface density profiles and cross-correlations, where the latter are reconstructed at an accuracy of 20\% down to scales of $k = 100\,h/$cMpc constituting a significant improvement over traditional approaches.
The presented method may become a key ingredient in future inference pipelines as it can be readily integrated into downstream analysis tasks of radio surveys.
\end{abstract}




\section{Introduction}\label{sec:introduction}
\begin{figure*}
    \includegraphics[width=\textwidth]{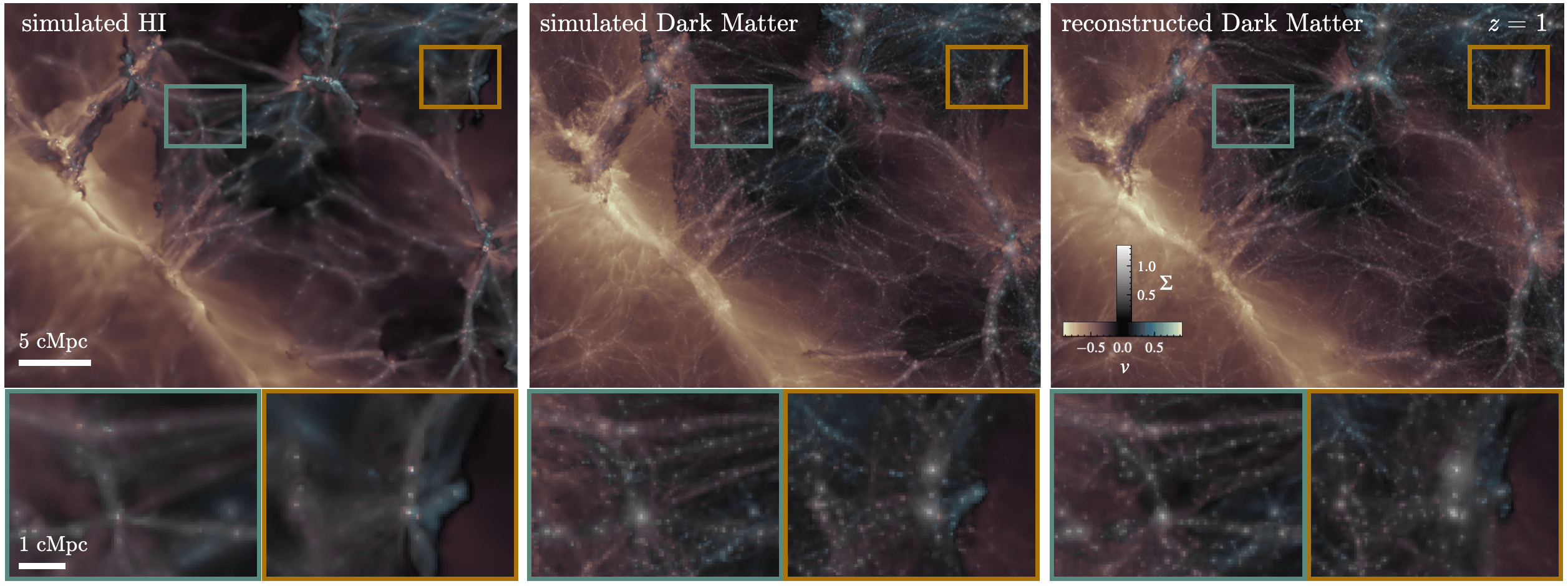}
    \caption{
    Overview figure showing an example region of the simulated HI and DM fields in the FB30 simulation and the corresponding reconstructed field from EMBER-2 at $z=1$. 
    The figure shows both normalized channels, surface density $\Sigma$ and line-of-sight velocity $v$, in a composite manner, where increased brightness indicates higher $\Sigma$, color-coded by the corresponding velocity of each pixel (from red to blue).
    For better visibility we show the normalized field values, which have $\mathcal{O}(1)$.
    This visual comparison highlights that the HI density and velocity fields contain significant amounts of information across different phase-space regimes, which can be used to accurately reconstruct the underlying DM distribution and kinematics from large down to small scale structures. 
    The two insets (teal and orange) showcase the model's reconstruction power on small scales for two example regions.
    }
    \label{fig:overview}
\end{figure*}
Dark matter is a fundamental ingredient of the $\Lambda$CDM model \citep{peebles_1984, carroll_2001}. 
While $\Lambda$CDM has been very successful 
in describing the large-scale structure and evolution of the Universe, the physical nature of dark matter, such as its particle mass and possible interactions, remains one of the most profound open questions in modern astrophysics \citep[e.g.][]{perivolaropoulos_2022}.
Weak Lensing (WL) has been established as a powerful tool to probe the total underlying matter distribution from galaxy tracers \citep[e.g.][]{bartelmann_2001}.
%
%
However, directly inferring dark matter properties from observational data remains challenging. 
In particular, WL requires large statistical samples \citep{takada_2014, perez_2020, mahony_2022, remy_2023}, and measuring the matter power spectrum requires accurate modeling of all systematic effects that contribute to the cosmic shear signal, such as baryonic feedback \citep[e.g.][]{Mohammed_2014, Schneider_2015, Schneider2019, Huang_2019, Schneider_2020, Lu_2021, Broxterman_2024}.

Complementary to luminous galaxies, HI provides another tracer of the dark matter distribution. 
The advent of next-generation radio surveys will supply an unprecedented wealth of data, establishing HI as a powerful probe for the link between dark and baryonic matter.
Pathfinder surveys such as MeerKAT's MIGHTEE \citep{maddox_21, heywood_2024, jarvis_2025}, and in the near future the Square Kilometre Array (SKA) telescopes \citep{smith_2015, Weltman2020}, are revolutionizing our ability to probe the distribution and kinematics of HI across a significant fraction of cosmic time.
While SKA-Mid will provide detailed observations of HI in the Interstellar Medium (ISM) and, to some extent, the Circumgalactic Medium (CGM) in the post-reionization Universe ($z \lesssim 3$) \citep[e.g.][]{coogan_2023}, observations via Intensity Mapping (IM) with SKA precursors already extend HI observations to the large-scale structure of the cosmic web at redshifts $z \lesssim 0.5$ \citep[e.g.][]{Cunnington_2023}.
%

In this letter, we present a novel Machine Learning (ML) model to predict the underlying dark matter distribution directly from HI tracers, providing a proof of concept that demonstrates the frameworks potential for future applications.
Traditional HI-based methods model HI as a biased tracer of dark matter, establishing the mapping via empirical relations often derived from numerical simulations.
For instance, Halo Occupation Distribution (HOD) models and abundance matching (AM) techniques construct a mapping between the mass of dark matter haloes and the contained HI mass \citep[][]{Berlind_2002, bagla_2010, Villaescusa_2014, Padmanabhan2016a, Padmanabhan2016b, Castorina_2017, wolz_2019, Qin_2022}.
Other methods model the relation as a bias expansion (BE) on the field level, where the HI and dark matter density fields are fitted with analytical descriptions with a few bias parameters \citep[][]{sarkar_2016, mcquinn_2018, modi_2019, qin_2022b, Obuljen_2023, Baradaran_2024, belsunce_2025}.
However, HOD and Bias Expansion approaches do not fully capture the complex interplay between HI and dark matter, especially on scales $<$ 1 Mpc \citep[e.g.][]{Obuljen_2023}, since the relation is established by either considering just a few quantities, such as the halo mass for HOD, or only a handful of fitting parameters in the BE, while discarding the remaining information stored in the field. 
Furthermore, the relation between HI and the underlying dark matter distribution is affected by feedback processes, such as stellar winds, ram pressure stripping and feedback from Active Galactic Nuclei (AGN), that redistribute baryonic matter on Mpc scales \citep[][]{angles_2017, hafen_2019, gebhardt_2024}.
These processes complicate dark matter inference from just a few simple HI properties, since environmental effects play a significant role in driving the scatter in the HI to dark matter halo mass relation. This behavior has been seen in observations \citep{Osterloo_2010, Janowiecki_2017, Cortese_2021, Li_2022, Saintonge_2022, Saraf_2024} and has been successfully reproduced in simulations \citep[e.g.][]{Rafieferantsoa_2015, Stevens_2019, Bernardini2022}.

ML methods have recently been applied to cosmological simulation datasets \citep{Villaescusa2021, Villaescusa2021b, rose_2025}, to learn the relation between dark matter and HI in a data-driven approach.
These works predict HI from the simulated dark matter counterparts using field level emulators \citep{Zhang2019, Zamudio2019, Thiele2020, Wadekar2021, Dai2021, andrianomena_2023, Sharma_2024, Bernardini2025, rigo_2025}.

Here we present a model that performs the "inverse" mapping, i.e. using simulated HI tracers to generate dark matter mock catalogs.
We adopt a data-driven approach to learn the direct mapping from HI to the underlying dark matter distribution for $z=0-6$, by taking into consideration the entire field level information at a spatial resolution of 43 ckpc. By leveraging neural networks trained on cosmological simulations, we construct a model that captures the complex, non-linear relationship between HI and dark matter while accounting for baryonic effects. 
In a related work, \cite{andrianomena_2023} have presented an approach based on invertible methods, however, only for a single redshift slice ($z=0$) and comparatively coarser spatial resolution ($\sim 147$ ckpc). 

Unlike statistical techniques that assume a fixed bias relation between baryons and dark matter, our approach allows for a more adaptive and accurate reconstruction of the dark matter field in different cosmic environments, including larger and smaller halos, filaments, and voids.
In particular, we leverage the recently introduced EMBER-2 model \citep[][]{Bernardini2025} to learn the mapping between projected HI and dark matter density and velocity information.
%
%
We present a visual summary of this approach in Figure \ref{fig:overview} which shows composite images of surface density and radial velocity information for the HI input field and the simulated and reconstructed dark matter fields. Also shown are two inset regions centered on two low density HI patches to highlight the large dynamic range between different regions in the cosmic web.

This letter is structured in the following way. In section \ref{sec:simulations} we present the simulations used for the training data. Section \ref{sec:ember2} briefly introduces the EMBER-2 framework and describes the training routine, while we discuss results in section \ref{sec:results} and conclude in section \ref{sec:conclusions}.

\section{Simulations}\label{sec:simulations}
Similar to the approach in \citet{Bernardini2025} we use cosmological hydrodynamical simulations that are part of the Feedback in Realistic Environments (\textsc{FIRE}\footnote{See the official \textsc{FIRE} project website: 
\href{https://fire.northwestern.edu}{\texttt{fire.northwestern.edu}}}) project.
The simulation used to create the training dataset is a volume simulation from the FIREbox suite \citep[][]{Feldmann2023, Feldmann2025} with a box length of $30\,\text{cMpc}/h$ (44 cMpc; hereafter denoted as \fbb{}, \citealt{Bernardini2025}). 
The simulation is run with {\sc gizmo} \citep{Hopkins2015}\footnote{A public version of \textsc{GIZMO} is available at \href{http://www.tapir.caltech.edu/~phopkins/Site/GIZMO.html}{\texttt{tapir.caltech.edu/phopkins/Site/GIZMO}}} using  the FIRE-2 galaxy formation physics model \citep[][]{Hopkins2018}.
The initial conditions for the simulation are created using MUSIC \citep[][]{Hahn2011}. The simulations are run with Planck 2015 cosmology \citep[][]{Planck2016}: $H_0$= 67.74 km/s/Mpc, $\rm \Omega_m$=$0.3089$, $\Omega_\Lambda$=$0.6911$, $\rm \Omega_b$=$0.0486$, $\sigma_8$=$0.8159$ and $n_\mathrm{s}$=$0.9667$. 
The FIRE-2 galaxy formation model has been validated in both zoom-in and cosmological simulations, reproducing cosmological HI statistics such as the total HI content in the Universe and the HI column density distribution function (CDDF) \citep[e.g.][]{Bernardini2022, Feldmann2023}, as well as properties on galactic scales, including HI mass fractions, disk scale heights, power spectra, and covering fractions \citep[e.g.][]{Gensior2023, Gensior2024, Tortora2024}.
The simulation is initialized with $1024^3$ dark matter and $1024^3$ gas particles with mass resolutions of $m_{\text{dm}}=2.68\times 10^6 \, \Msun$ and $m_{\text{gas}}=5 \times 10^5 \, \Msun$. For dark matter (star) particles the softening length is fixed to 160 (32) pc. Gas particles have a variable softening lengths with a minimum of 4 pc.
The most massive dark matter halo at $z = 0$ has a mass of $2.45 \times 10^{13} \, \Msun$, and the resolution of the simulation allows to resolve halos down to $\sim 10^9 \, \Msun$ (with $\geq 10^3$ particles).
Dark matter halos are identified with the Amiga Halo finder (AHF) \citep{Knollmann2009}.\footnote{See the official AHF website:  \href{http://popia.ft.uam.es/AHF/}{\texttt{popia.ft.uam.es/AHF/}}.}
In addition to \fbb{}, we use the FB512 simulation from the FIREbox simulation suite, hereafter denoted as \fb. This simulation is generated with a different random seed than \fbb{}, but uses the same physics and resolution, albeit in an 8 times smaller volume. In this work, \fb{} is only used as an independent test set to verify the models accuracy of the dark matter reconstruction.

\section{EMBER-2}\label{sec:ember2}
EMBER-2 is a neural network based model introduced in \cite{Bernardini2025}, designed to perform probabilistic mapping from dark matter to baryon fields across a large range of redshifts ($z=0-6$). In this letter we retrain the EMBER-2 model on the task of reconstructing the underlying dark matter fields from the simulated HI counterparts.
The task of our model is to perform the mapping from 2-dimensional projected HI density and radial velocity ($x$) to 2-dimensional dark matter density and radial velocity ($y$) for the redshift interval $z=0-6$, i.e. for each redshift the model learns the following probabilistic mapping
\begin{equation}
    f: (\Sigma_\mathrm{HI}, v_\mathrm{HI}) \rightarrow (\Sigma_\mathrm{dm}, v_\mathrm{dm}).
\end{equation}

EMBER-2 is a conditional GAN (cGAN) \citep{Goodfellow2014, Mirza2014} encompassing two neural networks. The training task of the generator $G$ consists of generating dark matter samples, $G(\eta|x, z)$, that are following the true data distribution $p_y$ as closely as possible. Here, $\eta$ is a noise input drawn from a normal distribution, $x$ the conditional HI map and $z$ the redshift.
Given $x$, the discriminator $D$ is tasked to distinguish between real ($y$) and generated samples ($G(\eta|x, z)$).
In this process $G$ learns the underlying conditional probability $p(y|x, \eta)$, which is inferred implicitly and thus intractable.
The loss functions are designed such that the two networks compete in an adversarial game, trying to outperform their respective opponent. We use the non-saturating GAN losses \citep{Goodfellow2014},
\begin{align}
\mathcal{L}_D &= - \mathbb{E}_{\eta} \left[ \ln \left(1 - D(G(\eta|x, z)\,|\,x, z) \right) \right] -\mathbb{E}_{y} \left[ \ln D(y|x, z) \right] \\
\mathcal{L}_G &= -\mathbb{E}_{\eta} \left[ \ln D(G(\eta|x, z)|x, z) \right]
\end{align}
where $y$ and $x$ are samples from the true data distributions $p_y$ and $p_x$. $\mathbb{E}$ is the expectation operator.
%
%
We follow the same approach as in \cite{Bernardini2025} to produce the datasets for training and testing the neural network model. Briefly, we use equidistant spacing of 0.25 between redshifts $6 \geq z \geq 2$ and a spacing of 0.1 between $2 \geq z \geq 0$. We deposit the particle information aggregated in slabs of $1.5$ Mpc$/h$ for HI and dark matter onto 2-dimensional grids of $1024^2$ pixels and a spatial pixel resolution of $\sim$43$\,\text{ckpc}$ using the \texttt{smooth} and \texttt{tipgrid} codes.\footnote{Code is publicly available on \href{https://github.com/N-BodyShop/smooth}{\texttt{github.com/N-BodyShop}}.}
During training we randomly crop patches of $128^2$ pixels from the projections and add random rotation and flipping as data augmentation.
We use a 3-fold cross-validation approach, where we take the projections from two axes for training while the remaining axis of \fbb{} is used for testing. Additionally, we complement the testing dataset with all axes from the \fb{} simulation.

We make use of the scaling functions introduced in \cite{Bernardini2025} to normalize the HI and dark matter surface density and velocity fields.
Furthermore we do not perform any additional hyper-parameter tuning and use the same parameter setup that was used to train the EMBER-2 model.\footnote{For additional details we refer the reader to the official project website 
\href{https://maurbe.github.io/ember2/}{\texttt{maurbe.github.io/ember2}}.}

\section{Results and Discussion}\label{sec:results}
%
%
\begin{figure}
    \includegraphics[width=\columnwidth]{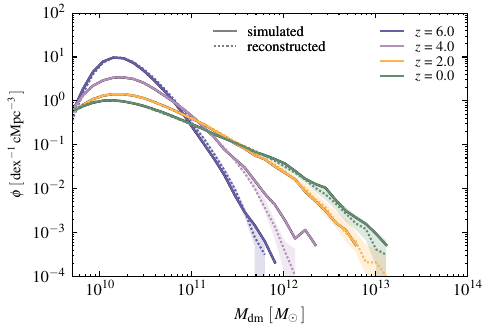}
    \caption{
    Structure mass functions $\phi$ for simulated and reconstructed dark matter maps at different redshifts.
    Lines and shaded bands indicate median and 16th to 84th percentiles for the test dataset.
    Across most redshift and structure-mass regimes, the reconstructed $\phi$ is in excellent agreement with simulations.
    }
    \label{fig:hmf}
\end{figure}


\begin{figure}
    \includegraphics[width=\columnwidth]{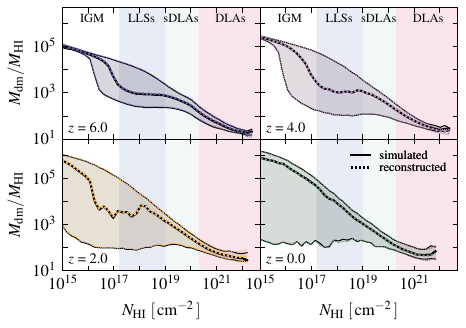}
    \caption{
    Dark matter to HI ratios as a function of peak $N_{\rm HI}$ and redshift (bottom left in each panel).
    Vertical shaded regions indicate different systems, showing the IGM regime, LLSs, sub-DLAs (sDLAs) and DLAs.
    Simulated (reconstructed) results are shown as solid (dashed) lines indicating the median and the 16th to 84th percentiles.
    The reconstructed median ratios and their scatter are in excellent agreement with the simulated counterparts.
    }
    \label{fig:f_dm}
\end{figure}

\begin{figure}
    \includegraphics[width=\columnwidth]{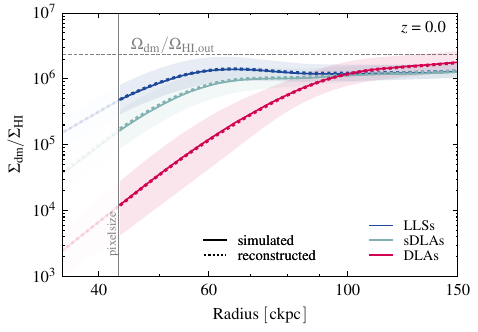}
    \caption{
    Simulated (solid) and reconstructed (dashed) dark matter surface density profiles as a function of radius at $z=0$.
    Colors correspond to the same $N_{\rm HI}$ thresholds as in figure \ref{fig:f_dm}.
    Lines indicate median relations whereas shaded regions show the 16th to 84th percentiles for the emulation. The vertical gray line indicates the pixel resolution of the maps, below which, values are interpolated.
    The dashed line represents the ratio of total dark matter to HI mass outside of structures.
    }
    \label{fig:profiles}
\end{figure}

\begin{figure}
    \includegraphics[width=0.95\columnwidth]{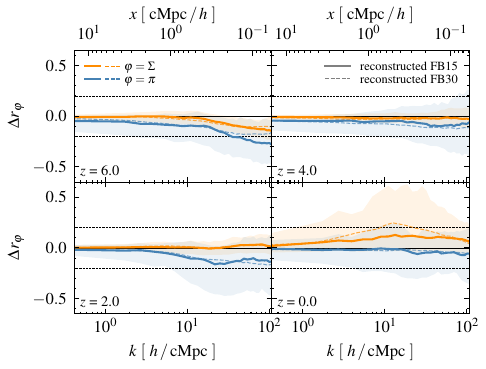}
    \caption{
    Median errors on the cross-correlation coefficients between the HI and dark matter surface densities ($\varphi = \Sigma$), as well as for the kinetic energy density surface density ($\varphi = \pi$) for selected redshifts.
    Solid and dashed lines represent the results on the FB15 and the FB30 test set, while for FB15 we also show the 16th to 84th percentiles as shaded regions. The horizontal dashed lines indicate the 20\% error band.
    }
    \label{fig:cross_corr}
\end{figure}

Figure \ref{fig:overview} presents a composite image displaying the two HI input channels (left), the simulated dark matter counterpart (middle), and the reconstructed result (right) for an example region at $z=1$. The two inset regions (highlighted in teal and orange) illustrate the performance of EMBER-2 across a wide dynamic range of densities and velocities, highlighting structural similarity between the simulation and the emulation.

We evaluate the neural network performance using a variety of summary statistics to probe field-level metrics and quantities derived from individual dark matter structures. Specifically, we analyze halo-based metrics including the mass function of emulated dark matter structures, dark matter to HI fractions and surface density profiles. We also perform a cross-correlation analysis to quantify the spectral statistics of the reconstructed fields.

For the simulation, halo catalogs can readily be computed using AHF and are used to compute halo-based metrics. However, we do not have access to such catalogs for the reconstructed maps.
Thus, we define the following algorithm based on the HI ground truth maps to analyze underlying dark matter structures that can be compared between simulation and emulation:\footnote{The entire algorithm is implemented using \href{https://scikit-image.org/}{\texttt{scikit-image}}.}
\begin{enumerate}
    \item
    To exclude background pixels, a threshold is applied to the HI density map where only pixels with values $\geq 10^{-3} \; \Msun / \rm ckpc^{2}$ are retained.
    \item Peaks in the thresholded map are identified using a local maxima finder. These mark the centers of HI and dark matter structures. We then measure several properties of the dark matter structures within a fixed circular aperture with fixed radius $R_{\rm fix}=150$ ckpc.
    \item For each structure we compute its total dark matter mass $M_{\rm dm}$, as well as its dark matter fraction as the ratio of dark matter and HI mass contained within its area $A=\pi R_{\rm fix}^2$. For each structure we also compute its peak HI number density $N_{\rm HI}$, defined as the highest $N_{\rm HI}$ pixel value within $A$, and use it to assign the structure to one of the following classes:
    Intergalactic Medium (IGM) if $N_{\rm HI} \leq 10 ^{17.2} \rm \, cm^{-2}$, Lyman Limit Systems (LLSs) if $N_{\rm HI} \leq 10^{19} \rm \, cm^{-2}$, sub-Damped Lyman-$\alpha$ absorbers (sDLAs) if $N_{\rm HI} \leq 10^{20.3} \rm \, cm^{-2}$ and DLAs if $N_{\rm HI} > 10^{20.3} \rm \, cm^{-2}$, where the threshold values are adopted from \citet{peroux_2020}.
    Using bilinear interpolation, we also extract radial surface density profiles $\Sigma_{\rm dm}(r)$ and $\Sigma_{\rm HI}(r)$ for concentric annuli around the structure center.
\end{enumerate}

We then apply this algorithm to the simulated and reconstructed dark matter maps and use the resulting structure catalogs in our further analysis.
\\

%
%
%

\noindent\textbf{Mass function:} From the catalogs we compute the structure mass function (SMF) $\phi \equiv \mathrm{d}n / \mathrm{d} \lg M_{\rm dm}$, i.e. the number of identified structures as a function of their dark matter mass $M_{\rm dm}$ as shown in Figure \ref{fig:hmf}.
%
The comparison between simulated and reconstructed maps shows that EMBER-2 reproduces realistic dark matter number statistics at all redshifts, particularly for $\lg M_{\rm dm} \leq 12$.\footnote{Unless otherwise stated, masses are in units of $\Msun$.} At higher masses, the comparison is limited by small-number statistics and Poisson noise.
\\

\noindent\textbf{HI fraction and radial profiles:} 
Given the identified structures, we also compare simulated and reconstructed dark matter mass fractions in Figure \ref{fig:f_dm}. The results show that the median dark matter to HI ratios as a function $N_{\rm HI}$ are very well reconstructed over the entire mass range, capturing the very HI deficient small structures in the IGM and LLSs, as well as higher mass systems reaching into the DLA regime. Furthermore, for all redshifts and mass regimes, EMBER-2 also accurately reconstructs the correct intrinsic scatter over the entire dynamic range. 
%
In Figure \ref{fig:profiles}, we show the radial profiles of $\Sigma_{\rm dm} / \Sigma_{\rm HI}$ at fixed redshift $z=0$ as a function of radius for LLSs, sDLAs and DLAs. For all three $N_{\rm HI}$ bins, the reconstructed profiles are in very good agreement with the simulation. In fact, EMBER-2 reconstructs the correct median amplitude and slope, both in the inner and outer part of structures, while for larger radii, the profiles converge towards the ratio of dark matter to HI outside of structures, denoted as $\Omega_{\rm dm} / \Omega_{\rm HI, out}$.
\\

\noindent\textbf{Cross-correlations:} After $z \lesssim 6$ the HI density distribution increasingly decouples from the underlying dark matter distribution. In fact, for $z \lesssim 2$ almost the entire HI mass resides within dark matter halos \citep[see e.g.][]{Villaescusa2018, Feldmann2023}. Hence, the scale at which the two fields decouple increases for lower redshifts, which is measured by the cross-correlation coefficient \cite[see e.g. Figure 2 in][]{Bernardini2025}.
For a specific property $\varphi$ (e.g. surface density), the cross-correlation coefficient is defined through the power spectra of dark matter ($P_{\rm dm, \varphi}$) and HI ($P_{\rm HI, \varphi}$) and their corresponding cross-power spectrum $P_{\rm dm+HI, \varphi}^{\times}$, as
\begin{equation}\label{eq:r}
    r_\varphi(k) = \frac{P_{\rm dm+HI, \varphi}^{\times}}{\sqrt{P_{\rm dm, \varphi} P_{\rm HI, \varphi}}}.
\end{equation}
We define the error on the cross-correlation coefficient between reconstructed and simulated maps as $\Delta r_\varphi =  r^{\rm rec}_{\varphi} / r^{\rm sim}_{\varphi} - 1$.
To quantify the correlations between individual dark matter channels, we also combine the two quantities $\Sigma$ and $v$ (both for HI and dark matter) to derive the kinetic surface energy density, $\pi = \Sigma \times v^2 / 2$.
In Figure \ref{fig:cross_corr} we show $\Delta r_\varphi$ at different redshifts by correlating $\Sigma_{\rm dm}$ with $\Sigma_{\rm HI}$, as well as $\pi_{\rm dm}$ with $\pi_{\rm HI}$.
The median errors for the \fb{} (\fbb{}) test sets are shown as solid (dashed) lines, while shaded regions represent the 16th to 84th percentiles for \fb{}.
The results show that EMBER-2 is capable of reconstructing the correct cross-correlations between those fields with errors less than $20\%$ for both $\Sigma$ and $\pi$, for most redshifts and scales up to $k=100 \; h/\rm cMpc$. This result holds both for the \fb{} and \fbb{} test sets.\footnote{Additional analysis on the bispectrum is presented in the online supplementary material.}
Traditional HOD (BE) methods typically achieve and accuracy of 10\% (50\%) for $k \lesssim 1\,h/$cMpc at $z=0$ \citep[see e.g. figure 3 in][]{Wadekar2021, Obuljen_2023}, highlighting the additional information stored in the field.
The only noticeable exception for \fbb{} occurs at $z=0$ where $\Delta r_\Sigma$ is slightly raised to $\sim 25\%$. We have investigated this behavior and found that it originates in the small number of massive halos in \fbb{}, while $\Delta r_\Sigma$ is disproportionally sensitive to the highest mass systems for large values of $k$ \citep[see Figure 10 in][]{Bernardini2025}.
In fact, for \fb{} the model reconstructs the cross-correlation coefficients at $z=0$ with deviations of maximum $\sim 20\%$ for all $k$-scales, motivating the conclusion that the deviation in \fbb{} is due to the small sample size of massive systems.

\section{Summary and outlook}\label{sec:conclusions}
In this letter we have used the EMBER-2 framework to learn the mapping from HI to dark matter on the field level over the redshift range $z=0-6$. 
Our analysis demonstrates that EMBER-2 can reconstruct realistic dark matter density and radial velocity fields from HI observations, with the correct number statistics, mass fractions and radial profiles over a wide range of masses and $N_{\rm HI}$ regimes.
Additionally, the cross-correlation analysis shows very good agreement between predicted and true dark matter density and kinematics, with maximum deviations for the dark matter surface density of $\sim 20\%$, up to scales of $k=100\,h/$cMpc.\\
\\
In the following we discuss the impact and outlook of the method for upcoming HI surveys and dark matter inference studies:
\begin{itemize}
    
    \item 
    Thanks to the fully convolutional architecture of EMBER-2, our method is well suited for dark matter inference pipelines across a broad range of scales: it captures the correlations between HI and dark matter on Mpc scales, while also reconstructing the dark matter to HI density profiles of individual halos. If needed, this adaptability allows the framework to be tuned and retrained to match the observational regime of interest, from large-scale HI surveys to small-scale probes of the HI distribution within halos.
    In general, we found that model inference becomes easier for larger spatial scales, due to the linear growth of structure as seen in Figure \ref{fig:cross_corr}.
    
    \item 
    One limitation of the current method is that it was trained on a fixed galaxy formation model.
    Our method naturally allows to incorporate simulations with multiple stellar and AGN feedback models, which we plan to incorporate in future work. 
    This will improve the methods's generalizability across different astrophysics models to marginalize over baryonic effects. 
    Since individual maps can be sampled on time-scales of $\sim$seconds, the presented method is fast enough to be incorporated in downstream analysis pipelines with realistic observations.

    \item To adapt to observational limitations, future iterations may also be retrained on degraded HI maps to match sensitivity limits of the SKA telescopes at different redshifts. In this pipeline the simulated HI fields are first degraded according to expected observational noise levels and resolutions, and subsequently used as input for EMBER-2 to reconstruct the underlying dark matter distribution.
    This extended framework is well-suited for existing analysis frameworks, such as KARABO\footnote{Code website can be found here: \href{https://i4ds.github.io/Karabo-Pipeline/}{\texttt{github.io/Karabo-Pipeline}}} \citep{Sharma_2025}, with its submodules OSKAR and RASCIL for simulating instrumental systematics and noise contamination. This integration may enable joint modeling between simulated and observational data facilitating parameter inference in realistic observational scenarios.
    
    \item Finally, expanding the input feature space of EMBER-2 is in principle straight-forward, and thus, additional channels, such as the stellar light measured from upcoming optical surveys (e.g. LSST, ELT) may be readily integrated. This multi-wavelength approach could likely enhance the accuracy of dark matter reconstructions by leveraging observational tracers complementary to HI.

\end{itemize}
\noindent These additions may broaden the applicability of the EMBER-2 framework to a wider range of physical models \citep[e.g.][]{Villaescusa2021b, rose_2025}, while also incorporating observational effects.
Paired with HI observations, such a model could readily be used together with sampling algorithms to perform parameter inference to constrain the physics of dark matter while marginalizing over baryonic effects.

\section*{Acknowledgments}
We thank the anonymous referee for his valuable comments that helped improve this work.
The \fb{} simulation was supported in part by computing allocations at the Swiss National Supercomputing Centre (CSCS) under project IDs s697, s698, and uzh18. 
The \fbb{}{} simulation was supported by computing allocations at CSCS under project IDs s1255 and uzh18.
MB and RF acknowledge financial support from the Swiss National Science Foundation (grant no 200021\_188552 and CRSII5\_193826).
DAA acknowledges support from NSF CAREER award AST-2442788, NASA grant ATP23-0156, STScI grants JWST-GO-01712.009-A, JWST-AR-04357.001-A, and JWST-AR-05366.005-A, an Alfred P. Sloan Research Fellowship, and Cottrell Scholar Award CS-CSA-2023-028 by the Research Corporation for Science Advancement.
JG gratefully acknowledges funding via STFC grant ST/Y001133/1.
This work made use of infrastructure services provided by \href{www.s3it.uzh.ch}{S3IT}, the Service and Support for Science IT team at the University of Zurich. 
The following software was used in this work: Numpy \citep{harris_2020}, Matplotlib \citep{hunter_2007}, scikit-learn \citep{pedregosa_2011} and PyTorch \citep{paszke_2019}.

\section*{Data Availability Statement}
The data used to produce the plots will be made available upon reasonable request to the corresponding author.


\bibliographystyle{mnras}
\bibliography{main}


\appendix
\section{Complementary analysis on the dark matter to HI mass}
\begin{figure}
    \centering
    \includegraphics[width=\columnwidth]{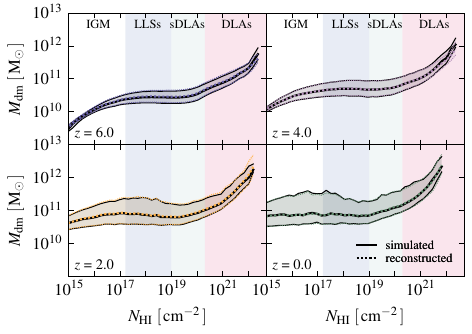}
    \caption{
    Similar to Figure \ref{fig:f_dm} but showing only the dark matter structure mass $M_{\rm dm}$ as a function of $N_{\rm HI}$.
    }
    \label{fig:NHI_Mdm}
\end{figure}

Complementary to Figure \ref{fig:f_dm}, we present in Figure \ref{fig:NHI_Mdm} the simulated and reconstructed dark matter structure masses without the additional $M_{\rm HI}$ normalization. This confirms that the network accurately recovers the median relation and scatter, independent of the additional variance introduced by the $M_{\rm HI} - N_{\rm HI}$ relation.

\section{Individual structure analysis}
\begin{figure}
    \centering
    \includegraphics[width=\columnwidth]{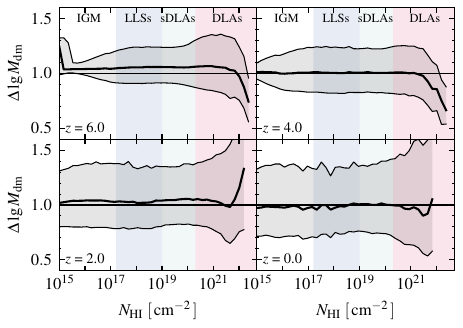}
    \caption{
    Logarithm of the ratio between the predicted and simulated dark matter structure masses as a function of $N_{\rm HI}$ and redshift.
    Shown are the median relation and the 16th to 84th percentiles.
    }
    \label{fig:delta_lgM}
\end{figure}

Figure \ref{fig:delta_lgM} shows the logarithmic ratio between reconstructed and simulated dark matter masses, $\Delta \lg M_{\rm dm} = \lg ( M_{\rm dm,\, rec} / M_{\rm dm,\, sim})$.
This complements the main-text analysis by comparing structures on an individual basis. The median accuracy is excellent across the full $N_{\rm HI}$ range, with an approximately constant scatter, while at $z=6$ we find a systematic offset of $\sim$ 0.05 dex and increasing scatter toward higher $N_{\rm HI}$.

\section{Bispectrum analysis}
\begin{figure}
    \centering
    \includegraphics[width=\columnwidth]{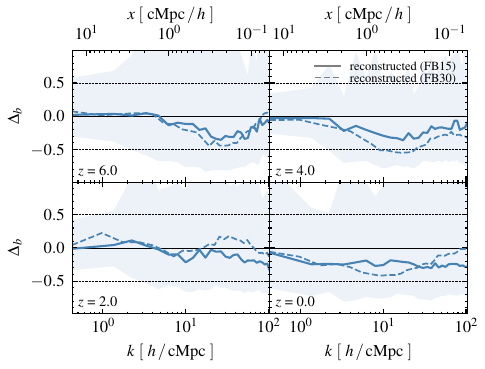}
    \caption{Fractional bispectrum errors between simulated and emulated dark matter surface density maps shown for both the \fb{} and \fbb{} simulations. Thick lines are medians, shaded regions indicate the 16th to 84th percentiles, and the vertical dashed black lines indicate the $50\%$ error bands.
    }
    \label{fig:bispectrum}
\end{figure}

In order to probe the agreement between reconstructed and simulated dark matter fields for higher order moments, we compute the error on the bispectrum $b$ of dark matter surface density fields defined as
\begin{equation}
    \Delta_b =  b^{\rm rec}_{\Sigma} / b^{\rm sim}_{\Sigma} - 1.
\end{equation}
The bispectrum of a 2-dimensional field is a function of two vector magnitudes $k_1$ and $k_2$ and an opening angle $\theta$, which we fix to $\pi / 2$ in this analysis.
The results are shown in Figure \ref{fig:bispectrum}, where we show medians (thick lines) and percentiles (shaded regions) of $\Delta_b$ as a function of $k$-vector at selected redshifts $z=0, 2, 4$ and 6. Overall, both for the FB15 and FB30 test sets, the error on the bispectrum of reconstructed fields is within 50$\%$ for most $k$-scales and redshifts.  
At higher redshifts ($z \geq 2$) the network is capable of reconstructing the correct bispectra up to spatial scales of $\sim 1$ cMpc$/h$ within $20\%$ errors, while for higher $k$ the reconstruction becomes more difficult. Potentially, a larger training dataset with varying initial conditions and larger boxsize might help reduce $\Delta_b$ even further.
We have done the same analysis for different opening angles and found that these conclusions are largely independent of $\theta$.
\end{document}